\documentclass[a4paper,12pt]{article}
\usepackage{amsmath}
\usepackage{amssymb}

\usepackage{color}
\usepackage[colorinlistoftodos]{todonotes}
\usepackage{color,amsmath}
\usepackage{pgf,tikz}
\usetikzlibrary{arrows}
\usetikzlibrary{decorations}
\usetikzlibrary{shapes,snakes,arrows}
\usepackage{multicol}
\usepackage{subfigure}
\usepackage{authblk}

\usetikzlibrary{automata}
\usetikzlibrary{positioning}
\usetikzlibrary{shapes.multipart}
\usepackage{rotating}

\usepackage{booktabs}
\usepackage{multirow}
\usepackage{lscape}

\newtheorem{definition}{Definition}
\newtheorem{proposition}{Proposition}
\newtheorem{example}{Example}\rm
\newcommand{\setC}{\mathcal{C}}

\begin{document}

\title{Novel Integer Programming models for the stable  kidney exchange problem \footnote{This work is financed by the ERDF – European Regional Development Fund through the Operational Programme for Competitiveness and Internationalization - COMPETE 2020 Programme, and by National Funds through the Portuguese funding agency, FCT -  Funda{\c{c}}{\~a}o para a Ci{\^e}ncia e a Tecnologia, within project ``mKEP - Models and optimization algorithms for multi­country kidney exchange programs" (POCI-01-0145-FEDER-016677), by COST Action CA15210 ENCKEP, supported by COST (European Cooperation in Science and Technology) -- http://www.cost.eu/. Bir\'o is supported by the Hungarian Academy of Sciences under its Momentum Programme (LP2016-3), and by the Hungarian Scientific Research Fund -- OTKA (no.\ K129086).}}


\author[1]{Xenia Klimentova}
\author[2,3]{P\'eter Bir\'o }
\author[1,4]{Ana Viana}
\author[1]{Virginia Costa}
\author[1,5]{Jo\~ao Pedro Pedroso}
\affil[1]{\small{INESC TEC, Campus da Faculdade de Engenharia da Universidade do Porto, Rua Dr. Roberto Frias, 4200-465, Porto,
Portugal}}
\affil[2]{Institute of Economics, CERS, T\'oth K\'alm\'an u. 4, 1097, Budapest, Hungary}
\affil[3]{Corvinus University of Budapest, F\H{o}v\'am t\'er 8, 1093, Budapest, Hungary}
\affil[4]{School of Engineering, Polytechnic of Porto, Rua Dr. Ant\'onio Bernardino de Almeida, 431, 4249-015 Porto
Portugal}
\affil[5]{Faculty of Sciences, University of Porto, Rua do Campo Alegre, 4169-007, Porto,  Portugal}


%
\maketitle
\begin{abstract}

Kidney exchange programs (KEP's) represent an additional possibility of transplant for patients suffering from end stage kidney disease. If a patient has a willing living donor with whom the patient is not compatible, the pair patient--donor can join a pool of incompatible pairs and, if compatibility between patient and donor in two our more pairs exists, organs can be exchanged between them. The problem can be modeled as an integer program that, in general, aims at finding the pairs that should be selected for transplant such that maximum number of transplants is performed.

In this paper we consider that for each patient there may exist a preference order over the organs that he/she can receive, since a patient may be compatible with several donors but may have a better fit over some than over others. Under this setting, the aim is to find the maximum cardinality stable exchange, a solution where no blocking cycle exists. For this purpose we propose three novel integer programming models based on the well-known edge and cycle formulations. These formulations are adjusted for both finding stable and strongly stable exchanges under strict preferences and for the case when ties in preferences may exist. Furthermore,
we study a situation when the stability requirement can be relaxed by addressing the trade-off between maximum cardinality versus number of blocking cycles allowed in a solution. The effectiveness of the proposed models is assessed through extensive computational experiments on a wide set of instances.

{\textbf{Keywords:} OR in health services\and stable exchange \and kidney exchange programs \and integer programming \and $k$-way exchange}
\end{abstract}
\renewcommand{\thefootnote}{\arabic{footnote}}

\section{Introduction}\label{sec:intro}
\subsection{Background}
End-stage kidney disease (ESKD) affects a considerable number of people and is a major threat to public health: the number of ESKD patients is expected to increase at an annual rate of 5 to 8\% in developed countries; more than 700,000 U.S citizens are currently receiving treatment for ESKD, the 9$^{th}$ leading cause of deaths each year in that country \cite{kidneyFound}.

Compared to dialysis, kidney transplant brings much better quality of life to the patient but the deceased kidneys available for transplants can only meet a tiny fraction of the existing demand. Living donation is an alternative that in the past was constrained to related living donors: if a patient had a related willing donor that was blood and tissue compatible with him/her, the transplant could be performed.  However, if they were not compatible, the transplant could not proceed. To overcome this issue, a new possibility of transplantation arose with Kidney Exchange Programs (KEPs): for a patient with a willing donor of one kidney but medically incompatible for transplant, the patient-donor pair can join many other such incompatible pairs in a pool for potential exchange. In its simplest form, with two incompatible pairs, KEP facilitates kidney exchange if the donor in one pair is compatible with the patient in the second pair and vice-versa. The concept behind this exchange in a 2-cycle extends to $k$-cycle when $k$ incompatible pairs are involved. Within a KEP, a limit $K$ on the maximum length of exchange cycles is usually imposed, i.e., in the \emph{kidney exchange problem} only $k$-cycles with length $k\leq K$ are considered.

A relevant practical extension of the problem covers the case in which a donor with no associated patient is allowed in the program. That donor, hereby referred to as non-directed donor (NDD), initiates a \emph{chain} by donating his/her kidney to a patient in one pair. The donor in that pair donates to a patient in a subsequent pair and so forth. The last donor in the chain either donates to the deceased donor's waiting list or acts as a ``bridge''  donor in future matches. In many European programs a limit $L$ on the maximal length of a chain is also imposed.
Considering a dummy compatibility from all the pairs to each NDD,  chains can also be referred to as cycles, taking into account that their length is limited by $L$.

The underlying combinatorial optimization problem associated with a KEP can be modeled as a vertex-disjoint cycle packing problem in a digraph, having vertices as incompatible pairs or NDD donors and arcs indicating compatibility of donor in one pair with patient in another pair. The aim is in general to obtain the set of disjoint cycles that maximizes the number of associated transplants. Additional objectives that may encapsulate preferences by way of weights or hierarchically are, e.g. difference in age between patient and donor and HLA-matching between them \cite{Massie2016,Ashby2017}.

In this paper we assume that patients have preferences over potential donors, represented by ranks on in-going arcs, and aim at finding the maximum cardinality \emph{stable exchange}. A \emph{stable exchange} is a set of vertex-disjoint cycles such that there is no so-called \emph{blocking cycle} where all the vertices would be better off, according to their preferences, than in the current solution. If  strict preference is required only for one vertex in the blocking cycle then we speak about \emph{strongly stable exchanges}. Stability and strong stability are equivalent to the core and strong core properties, respectively, which are standard solution concepts in cooperative game theory, see \cite{SS1974} and \cite{RP1977}. The usage of stable matchings in two-sided matching markets are standard both in theory and practice, in applications such as resident allocation, college admission and school choice, see e.g.\ a recent book on the algorithmic aspects of this topic by~\cite{Manlove2013book}.  

When the length of the exchange cycles is not limited there is always a stable solution that can be found efficiently by Gale's Top Trading Cycle (TTC) algorithm \cite{SS1974}. Moreover, if the preferences are strict then the TTC solution is unique and strongly stable \cite{RP1977}, as suggested in the seminal paper on KEPs by \cite{RSU2004}. However, when the length of   cycles is limited, which is typically the case for KEPs, a stable or strongly stable solution may not exist and the problem of deciding its existence is NP-hard \cite{Biro2010,Huang2010} even for tripartite graphs (also known as the cyclic 3D stable matching problem \cite{NG1991}).

To the best of our knowledge, the concept of stable solutions has not been used yet in real KEPs. In fact, besides considering some logistical and quality factors, the typical objective in European programs is to maximize the number of transplants \cite{Biroetal2019a,Biroetal2019b}. However, there are several reasons why stable solutions may well be aimed for in the future. One reason is that quality of the transplants is becoming more important for the patients and system organizers as well, since increasing the expected graft survival time yields a longer expected life time for the patient and reduces the risk that this patient will move back to the waiting list for another transplant, in a state when it becomes more complicated to find a compatible donor. Secondly, there are more and more platforms where the patient-donor pairs can communicate, and so the possibility of finding a better exchange for some pairs may become feasible through alternative channels.
Finally, there is an increasing number of compatible pairs or ABO-incompatible pairs, who are joining the KEP pools instead of direct transplantation with the incentive of getting a better donor. In the UK, for example, this strict improvement can be guaranteed by the possibility of setting quality thresholds, which is an available option for every pair. A pair might strategically set its threshold high enough in the first matching run it participates in and then they gradually decrease it in later rounds if no exchanges are offered. Indeed, the actual solution can be close to being stable in such a dynamic system, even if the main optimization goal is to maximize the number of transplants. However, as this process takes time to converge, the health condition of the patients gets worse while waiting for convergence. A direct mechanism for finding stable exchanges can give the incentive to the pairs of reporting low thresholds, as they are guaranteed to get the best possible donor in a stable solution.
In fact, if no length limit is imposed then the unique strongly stable solution, obtained by the TTC algorithm satisfies the \emph{respecting improvement property}, which means that bringing a better quality (e.g.\ younger) or more compatible (e.g.\ blood-type O) donor can only improve the quality of the exchange donor that this patient will receive \cite{RIP2020}. This property is violated for the standard objectives of maximizing the size or the total weight (e.g., quality score) of the solution, which can be a serious concern.

\subsection{Our contribution}
In this paper we propose three novel Integer Programming (IP) models for the stable exchange problem that are suitable for finding  stable and strongly stable exchanges, considering both strict preferences and the possibility of ties. {It extends the work in \cite{Cos2018}
by providing more general and tighter formulations. Namely, the new models now allow for the incorporation of non-direct donors, permitting chains and cycles maximum length to be different. They also consider the possibility of ties on preferences.}
Furthermore, we propose a model where the stability requirement is relaxed and substituted by the minimization of the number of blocking cycles. For that model we evaluate the trade-off between that objective and the maximization of the number of transplants by imposing a constraint on the maximum number of transplants that can be sacrificed in order to obtain ``the most stable" solution, i.e. the solution with the least number of blocking cycles. Finally, we assess the performance of the proposed IP formulations on generated KEP instances, and we analyze the trade-off between maximality and stability of the solutions.

{To the best of our knowledge there is no previous work in the literature where it is  proposed  to compute stable matching in this setting with optimization techniques.} Therefore, the contribution of this paper to the state of the art is twofold: it enriches the scarce literature on the use of IP models to find stable matches; and it provides, for the first time, integer programming models for finding maximum cardinality stable exchanges in a KEP.

\subsection{Related literature}

Although kidney exchange programs represent a relatively new paradigm, considerable relevant research has been done during the last decades on the associated optimization problems. Research lines attacked the modeling aspects of the base problem and its variants, and  analyzed  different policies.

Starting with the proposal by \cite{Rapaport1986}, the medical literature contains a number of detailed descriptions on operating KEPs (e.g.  a survey by \cite{Glorie2014}, or the recent study by \cite{Agarwal2019} for the US). Thorough review of European KEPs has also been made recently and published as two deliverables of ENCKEP (European Network for Collaboration on Kidney Exchange Programs) COST Action:  the general operation of programs is summarized in \cite{Biroetal2019a} and the optimization aspects in \cite{Biroetal2019b}. 
Well-established quality estimators of the quality of living donations were recently proposed in the medical literature in \cite{Massie2016} and \cite{Ashby2017}.

Regarding the optimization aspects, seminal work on IP models for KEP's is presented in \cite{Abraham2007} and \cite{Roth2007}. The authors propose two models -- the edge and the cycle formulations -- none of the models being compact (i.e., none having both polynomial number of constraints and variables). Later, \cite{Constantino2013} proposed compact formulations for the basic KEP problem. The authors also showed how their model could be extended in order to accommodate the following problem variants: inclusion of non-directed donors, inclusion of compatible pairs and possibility of having more than one donor associated to a patient. Although they prove that linear relaxations of the compact formulations do not provide better upper bounds for optimal solutions, when compared to the cycle formulation, computational results reinforce the idea that compact formulations are of practical relevance:  for larger values of $K$ and especially if graphs are denser, compact formulations provide better results and are able to solve larger problems. More recently \cite{Dickerson2016} presented two new compact IP models and showed that one of those models has a linear programming relaxation that is as tight as the previously known tightest formulation -- the cycle formulation. The case of bounded length cycles and unbounded length (aka never ending) chains was studied from an optimization aspect in \cite{Anderson2015pnas}, and was successfully implemented in a national US KEP, Alliance for Paired Donation (APD), as described in \cite{anderson2015interfaces}. For recent surveys on the optimization aspects of KEPs, we refer to \cite{Mak2017} and \cite{AshlagiRoth2020}; a review on KEP simulations can be found in \cite{Santos17}.

The model of stable exchanges with unbounded length, which is equivalent to the classical housing market model by \cite{SS1974}, has been proposed as a possible solution concept for kidney exchange in the seminal paper by \cite{RSU2004}. 
However, when it became apparent that in the real applications the length of the cycles is limited and that the main objective is rather the maximization of the number of transplants, the literature started to focus on that sort of models~\cite{Roth2005}.
The only research papers we are aware of about bounded length stable exchanges motivated by the kidney exchange applications are \cite{Biro2010} and \cite{Huang2010}, but these papers only considered the computational complexity (i.e., NP-hardness) of the corresponding problem.

There is a recent line of research on computing stable matchings through integer programming methods for two-sided matching markets. That is the case for the hospital--resident problem with couples \cite{BMMcB2014}, ties \cite{KM2014,Delormeetal2019}, college admissions with special features \cite{ABMcB2016}, stable project allocation under distributional constraints \cite{ABSz2018}, and car--sharing~\cite{Wang2018}. For the algorithmic aspects of matching problems under preferences we refer to the book by \cite{Manlove2013book}.

\subsection{Layout of the paper}

The remaining of the document is structured as follows. In Section \ref{sec:ip} we describe our integer programming formulations for finding stable  and strongly stable exchanges. We continue by providing further IP formulations for the multiple objective case in Section \ref{sec:nearly_stable}, where the relaxed version of stability versus maximization of cardinality of exchange is analysed. Computational experiments are presented and discussed in Section \ref{sec:comp}.  In Section~\ref{sec:conclusion} we provide concluding remarks.

\section{Definitions and notation}\label{sec:defs}

Consider a digraph $G=(V,A)$, where $V=\{1,2, \dots n\}$ is the set of vertices and $A$ is the set of arcs. In the context of a KEP, vertices represent  patient-donor pairs or NDDs and an arc $(i,j)$ indicates  possibility of transplanting a kidney from donor in $i$ to a compatible patient in $j$. In case compatible pairs are included into the programme, for such a pair $j$  we may restrict the incoming arcs to those donors that are strictly better in quality than the donor associated to patient $j$. If $j$ is a NDD  we create  dummy arcs to $j$ from all other vertices that are not NDDs  (these arcs represent  donation of the last living donor in a chain to the deceased donors waiting list).

Let $\setC$ be the set of cycles in $G$ of length at most $K$, if all vertices represent patient-donor pairs, and of length at most $L$, if the cycle contains exactly one NDD. Denote by $V(c)$ and $A(c)$ the set of vertices and arcs, respectively, that are involved in $c\in\setC$.
Within this context, an \emph{exchange} is a set of vertex disjoint cycles $M\subseteq \setC$.
We say that  vertex $i$ is matched if there is a cycle  $c\in M$  such that $i \in V(c)$ and denote by $A(M) = \bigcup_{c\in M}A(c)$ the set of arcs belonging to exchange $M$.


\subsection*{Preferences}

For each vertex $j\in V$, each vertex $i$ from  set $\delta(j) = \{i~|~(i,j)\in A\}\subseteq V$ is ranked with value $r\in\{1,\dots,|\delta(j)|\}$. The \emph{preferences} of a vertex $j$ over vertices of  set $\delta(j)$ are defined as follows.

\begin{definition}\rm\label{def_pref} For $i,i^\prime\in\delta(j)$ ranked with $r,r^\prime$, respectively, vertex $j$ \emph{prefers} $i$ to $i^\prime$ (denoted by $i <_j i'$), if $r< r^\prime $.
\end{definition}

\begin{definition}\rm\label{def_indif} For $i,i^\prime\in\delta(j)$ ranked with $r,r^\prime$, respectively, vertex $j$ \emph{is indifferent} between $i$ and $i^\prime$ (denoted by $i =_j i^\prime$), if $r^\prime = r$.
\end{definition}

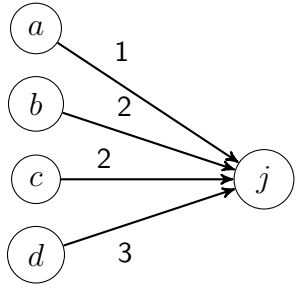
\begin{figure}[h]
 \begin{center}
 \begin{tikzpicture}[->, >=stealth', node distance = 3cm,scale = 0.2]
 \node[circle, draw] (b) at (20,0) {$b$};
 \node[circle, draw] (a) [above of = b, node distance = 1cm] {$a$};
 \node[circle, draw] (c) [below of = b, node distance = 1cm] {$c$};
 \node[circle, draw] (d) [below of = c, node distance = 1cm] {$d$};
 \node[circle, draw](i) [right of = c] {$j$};
\path[every node/.style={font=\sffamily\small}]
    (a) edge [thick] node[auto,near start] {1} (i)
    (b) edge [thick] node[auto,near start] {2} (i)
    (c) edge [thick] node[auto,near start] {2} (i)
    (d) edge [thick] node[auto,swap,near start] {3} (i);
 \end{tikzpicture}
 \caption{Example of preference or indifference of one vertex over others.}
 \label{fig:KEP_pref}
 \end{center}
 \end{figure}

Figure \ref{fig:KEP_pref} illustrates definitions \ref{def_pref} and \ref{def_indif}. Based on the definitions $a <_j b$, $a <_j c$, $a <_j d$, $b =_j c$, $b <_j d$ and $c <_j d$.

\begin{definition}\rm\label{def_strict}If $i=_jk$ implies that $i=k$ for all $j\in V$,  preferences are called \emph{strict}. Otherwise they are called \emph{weak}.\end{definition}

In this work we assume that a vertex always prefers to be matched (i.e., to belong to a cycle in the exchange), rather than be unmatched.
For the case of NDDs, to avoid harming of patients in the waiting list (who, in general are the recipients of  the kidney from the last donor in the chain), we consider that preferences are based on quality factors associated to the last donor in the chain: donors with higher quality measures are ranked higher by the dummy patient of the NDD. 



%

The notion of preferences for vertices can be extended for cycles.

\begin{definition}\rm\label{def_pref_cyc} Vertex $j$ \emph{prefers} cycle $c\in\setC(i)$ over cycle $c^\prime\in\setC(i)$, denoted as $c\prec_{j} c^{\prime}$, if for $(i,j) \in A(c)$  and $(i^\prime,j)\in A(c^\prime)$, $i<_ji^\prime$.
\end{definition}

\begin{definition}\rm\label{def_indif_cyc} Vertex $i$ is \emph{indifferent} between cycles $c$ and $c^\prime$ ($c\sim_ic^\prime$) if for $(i,j)\in A(c)$ and  $(i^\prime,j)\in A(c^\prime)$, $i=_ji^\prime$
\end{definition}
Note that for the case of strict preference the indifference between cycles reduces to the cycles that share an arc, i.e. $(i,j)\in A(c)\cap A(c^\prime)$.

\begin{definition}\rm\label{def_weak_pref_cyc}  Vertex $i$ \emph{weakly prefers} cycle $c$ to $c^\prime$ ($c\preceq_i c^\prime$) if it prefers $c$ to $c^\prime$ or it is indifferent between them.
\end{definition}



\subsection*{Stability and strong stability}

The \emph{Stable Exchange Problem} is the problem of finding a maximum cardinality \emph{stable or strongly stable exchange}, defined as follows~\cite{Biro2010,Huang2010}:

\begin{definition}\rm\label{def_stability}
 An exchange $M$ is called \emph{stable} if there is no blocking cycle $c \notin M$. A \emph{blocking cycle} $c \notin M$ is a cycle such that every vertex $i$ in $V(c)$ is either unmatched in $M$ or prefers $c$ to $c^\prime$, where $c^\prime \in M$ and $i\in V(c^\prime)$.
\end{definition}
\begin{definition}\rm\label{def_strong_stability}
 An exchange $M$ is called \emph{strongly stable}  if there is no weakly blocking cycle $c \notin M$. A \emph{weakly blocking cycle} is a cycle $c \notin M$ such that for every $i\in V(c)$, $i$ is either unmatched in $M$ or weakly prefers $c$ to $c^\prime \in M$, 
 with strict preference for at least one vertex.
\end{definition}
%

Although Definitions~\ref{def_stability} and~\ref{def_strong_stability} are the classic definitions for (strong) stability  we will consider their alternative form, 
also used in~\cite{Kar17}, as it is writes more natural when designing IP formulations.

\begin{proposition}
\label{prop_stability}
An exchange $M$ is \emph{stable} if and only if for every $c\in \setC$  there exists an arc $(i,j)\in A(c)$ and $(k,j)\in A(M)$ such that $j$ weakly prefers $k$ to $i$.
\end{proposition}

For the case of strong stability and strict preferences the following holds.
\begin{proposition}
\label{prop_strong_stability_strict}
In case of strict preferences an exchange $M$ is \emph{strongly stable} if and only if for every cycle $c\in \setC$, $c\notin M$, there exists an arc $(i,j)\in A(c)$ such that there exists $(k,j)\in M$ and $j$ prefers $k$ to $i$.
\end{proposition}

Although Proposition \ref{prop_stability} holds both for strict and weak preferences, Proposition \ref{prop_strong_stability_strict} is not valid for the weak preferences in case of ties.

\begin{example}\label{ex:ties}
 Figure \ref{fig:example_SS} exemplifies one case where proposition \ref{prop_strong_stability_strict} would fail.
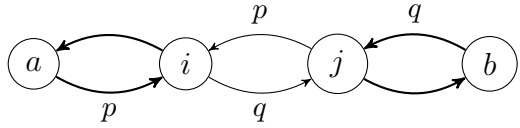
\begin{figure}[!htbp]
\centering
\begin{tikzpicture}[->, >=stealth', node distance = 2cm]
\node[circle, draw](i) at (0,0) {$i$};
\node[circle, draw](j) [right of=i] {$j$};
\node[circle, draw] (a) [left of = i] {$a$};
\node[circle, draw] (b) [right of = j] {$b$};

\path[every node/.style={font=\sffamily\small}]
    (j) edge [bend right] node[auto,swap] {$p$}  (i)
    (i) edge [bend right] node[auto,swap] {$q$} (j)
    (a) edge [bend right, thick] node[auto,swap] {$p$} (i)
    (b) edge [bend right, thick] node[auto,swap] {$q$} (j)
    (j) edge [bend right, thick] node {} (b)
    (i) edge [bend right, thick] node {} (a);
\end{tikzpicture}
 \caption{Strong stability with ties}
  \label{fig:example_SS}
\end{figure}
Assume that $i$ has equal preferences ($p$), over $j$ and $a$, and that $j$ also has equal preferences ($q$), over $i$ and $b$. Under this setting, consider an exchange $M$ that consist of cycles $(a,i)$ and $(j,b)$. Then for cycle $c = (i, j) \notin M$  there is no arc such that a vertex in $c$ strictly prefers its in-neighbour in $M$ to an arc in $c$ (as required by Proposition~\ref{prop_strong_stability_strict}). Nevertheless  cycle $c$ is not a blocking cycle accordingly to Definition~\ref{def_strong_stability}.
\end{example}

To accommodate this case, Proposition~\ref{prop_strong_stability_strict} is adapted as follows:
\begin{proposition}
\label{prop_strong_stability_ties}
In case of weak preferences, an exchange $M$ is \emph{strongly stable} if and only if for every cycle $c\in \setC$, $c\notin M$, either for all $(i,j)\in A(c)$ there exists an arc $(k,j) \in M$ such that $k=_ji$ (i.e. each vertex $j$ in $c$ is matched in $M$ with equal preference to in-neighbour $i$), or there exists an arc $(i,j)\in A(c)$ and an arc $(k,j)\in A(M)$ such that $j$ prefers $k$ to $i$.
\end{proposition}

Within Proposition~\ref{prop_strong_stability_ties}, cycle $(i,j)$ in Example~\ref{ex:ties} satisfies the first condition of the proposition.

\section{Integer programming models for stable exchanges}\label{sec:ip}
In this section we introduce three new IP models for the stable exchange problem defined above. The two first models originate directly from the edge and cycle formulations, while the third model incorporates the decision variables from the two previous ones. For each formulation we start by presenting the model for the case of no ties (i.e. when preferences are strict). Then, and whenever required, we introduce the necessary changes to accommodate ties.
%

\subsection{Edge Formulation} \label{sec:edge_form}

Consider a variable $y_{ij}$ associated with each arc $(i,j)\in A$ in the graph $G(V,A)$ and defined  in~\cite{Abraham2007} and~\cite{Roth2007}:

$$y_{ij} = \left\{ \begin{array}{lcl}
			1 & & \mbox{if arc } (i,j) \mbox{ is selected in the exchange, i.e }  (i,j)\in A(M), \\
			0 & & \mbox{otherwise.}
		\end{array}\right. $$


Consider also variables $y^c_{ij}$ $y^n_{ij}$, where:

\begin{itemize}

\item $y^c_{ij} = 1$, if arc $(i,j)$ is selected and is part of an exchange cycle of patient-donor pairs; 0, otherwise.

\item $y^n_{ij} = 1$, if arc $(i,j)$ is selected and is part of a NDD chain; 0, otherwise.
\end{itemize}

These extra variables extend the model in~\cite{Abraham2007} and~\cite{Roth2007}, allowing $K$ and $L$ to be assigned different values.

{Let $\mathcal{P}^c$ be the set of all non-cyclic paths $p$ in $G$ with $K$ arcs, with all vertices representing patient-donor pairs. In a similar way, let $\mathcal{P}^n$ be the set of paths with $L$ arcs, with the first vertex in a path representing a NDD and all other vertices being patient-donor pairs. Denote by  $A(p)$ the set of arcs in $p$.}

The integer program for finding a stable exchange is written as the edge formulation in~\cite{Abraham2007} and~\cite{Roth2007} with an additional set of constraints that enforce stability:
{
\begin{align}
& \text{Maximize}  & &  \sum_{(i,j)\in A} y_{ij}   \label{eq:EdgeObj}\\
& \text{Subject to:}     & &  \sum_{j:(i,j)\in A} y_{ij} 	  						 \leqslant   1 				& &\forall i\in V \label{eq:EdgeMostOne} \\
&               & & y_{ij} = y^c_{ij}+y^n_{ij} && \forall (i,j)\in A \label{eq:Arcconnect}\\
&               & &  \sum_{j:(j,i)\in A} y^c_{ji}  						 =		   \sum_{j:(i,j)\in A} y^c_{ij}	& & \forall i\in V, \label{eq:EdgeCInOut} \\
&               & &  \sum_{j:(j,i)\in A} y^n_{ji}  						 =		   \sum_{j:(i,j)\in A} y^n_{ij}	& & \forall i\in V, \label{eq:EdgeNInOut} \\
&			     & &  \sum_{(i,j) \in A(p)} y^c_{ij} 	 \leqslant  K-1 						& & \forall p\in\mathcal{P}^c \label{eq:EdgePathC} \\
&			     & &  \sum_{(i,j) \in A(p)} y^n_{ij} 	 \leqslant  L-1 						& & \forall p\in\mathcal{P}^n \label{eq:EdgePathN} \\
&    & &\sum_{(i,j)\in A(c)}\sum_{k:k \leq_j i} y_{kj} \geq 1, && \forall c\in\mathcal{C}.\label{eq:stab_edge}\\
&			     & &  y_{ij} \in \{0,1\}, y^c_{ij}\geq 0 ,y^n_{ij}\geq 0						\label{eq:Edge01}								& & \forall (i,j)\in A.
\end{align}

 The objective function~\eqref{eq:EdgeObj} maximizes the  cardinality of the exchange (in the context of a KEP this corresponds to maximizing the total number of transplants). Constraints~\eqref{eq:EdgeMostOne} guarantee that the cycles in the exchange are disjoint, i.e. a donor can only donate one kidney. Constraints~\eqref{eq:Arcconnect}, together with constraints~\eqref{eq:EdgeMostOne}, state that an arc cannot be simultaneously in a  patient-donor pair cycle and in a NDD chain. 
 Constraints~\eqref{eq:EdgeCInOut} and~\eqref{eq:EdgeNInOut} are standard flow conservation constraints that ensure that the number of kidneys received by the patient in pair $i$ is equal to the number of kidneys given by the associated donor.  Constraints~\eqref{eq:EdgePathC} enforce maximum cycle-length: to exclude cycles larger than $K$, every path of length $K$ arcs cannot have more than $K-1$ arcs in a feasible exchange. In a similar way, constraints~\eqref{eq:EdgePathN} enforce the maximum length of NDD chains.  These two sets of constraints require all paths of length $K$ and $L$ to be considered explicitly in the model and, in general, the number of such paths grows exponentially with $K$ and $L$.
Finally, constraints~\eqref{eq:stab_edge} ensure  stability: according to Proposition~\ref{prop_stability}, each arc $(i,j)$ in any cycle $c$ is either in the set of arcs of the exchange, or vertex $j$  is matched with a vertex that it prefers to $i$.
}

Recall that, for the case of stability,  Proposition~\ref{prop_stability} is valid both for strict and weak preferences. Differently, for the case of strong stability, Propositions~\ref{prop_strong_stability_strict} and~\ref{prop_strong_stability_ties} hold for the cases of strict and weak preferences, respectively. Accordingly, for finding a strongly stable exchange in the case of strict preferences constraints~(\ref{eq:stab_edge}) in model~\eqref{eq:EdgeObj}-\eqref{eq:Edge01} must be replaced by:

\begin{align}
\sum_{(i,j)\in A(c)} y_{ij} + |c|\cdot\left[ \sum_{(i,j)\in A(c)}\,\, \sum_{k:k<_j i} y_{kj}\right]   \geq |c|, && \forall c\in\mathcal{C}.\label{eq:s_stab_edge}
\end{align}

Following  Proposition~\ref{prop_strong_stability_strict}, constraints (\ref{eq:s_stab_edge}) ensure that each cycle  $c\in\mathcal{C}$ is either in an exchange (i.e. all its arcs are chosen) or there is at least one patient in $c$ that is matched with a strictly better donor.

Finally, for the case of weak preferences, constraints~(\ref{eq:stab_edge}) are replaced by:
\begin{align}
\sum_{(i,j)\in A(c)} (y_{ij}+\sum_{k: k=_j i, k\neq i} y_{kj})+|c|\cdot\left[ \sum_{(i,j)\in A(c)}\,\, \sum_{k:k<_j i} y_{kj}\right]   \geq |c|, && \forall c\in\mathcal{C}.\label{eq:s_stab_edge_ties}
\end{align}

These constraints are similar to~\eqref{eq:s_stab_edge} but the first term of  (\ref{eq:s_stab_edge_ties})  covers, in addition, the case where all patients in the cycle are matched with donors that are equally preferable to the ones in the exchange. The second term of the constraint remains the same as in~(\ref{eq:s_stab_edge}).

\subsection{Cycle Formulation}

Define variables $x_c$ for each cycle  $c \in \mathcal{C}$ as:

$$x_c = 	\left\{ \begin{array}{lcl}
			1 & & \mbox{if cycle $c$ is selected in the exchange,}  \\
			0 & & \mbox{otherwise.}
		       \end{array}
		\right. $$
For each pair $(i,c)$, $i\in V$, $c\in \setC(i)$, define also set $B_{i,c} = \{\bar c \in \setC(i), \bar c\neq c : \bar c \preceq_i c\}$, i.e. the set of cycles that are different from $c$ and weakly preferred by $i$ to $c$ (see Definition~\ref{def_weak_pref_cyc}). 
%
%

Using the cycle formulation, the stable exchange problem is modelled as follows.

\begin{align}
& & \text{Maximise}  &&& \sum_{c:c\in\mathcal{C}}\,|c|\cdot x_c. \label{eq:objcycle}\\
& & 	&&& \sum_{c:i\in V(c)} x_c \leq 1 			&&  \forall i \in V \label{eq:CycleMostOne} \\
&& &&& x_c+ \sum_{s\in\bigcup_{i\in V(c)}  B(i,c)} x_s\geq 1, && \forall c\in \setC, \label{eq:CyclePref}\\
& &			   &	& & x_c \in \{0,1\}					&& \forall c \in \mathcal{C}, \label{eq:Cycle01}
\end{align}

The objective function (\ref{eq:objcycle}) maximises the cardinality of the exchange. Constraints (\ref{eq:CycleMostOne}) guarantee that an exchange is a set of disjoint cycles, and constraints (\ref{eq:CyclePref}) guarantee stability: either $c$ is in the exchange or, for at least one vertex $i\in V(c)$, there exists another cycle $s$ weakly preferred by $i$.

For the case of strong stability, for each pair $(i,c)$, $i\in V$, $c\in \setC(i)$,  define set  $S_{i,c} = \{\bar c \in \setC(i) : \bar c \prec_i c\}$, i.e. the set of cycles that $i$ prefers to $c$ (see Definition~\ref{def_pref_cyc}). A strongly stable maximum cardinality exchange with strict preferences can be obtained by replacing constraints~\eqref{eq:CyclePref} in problem~\eqref{eq:objcycle}-\eqref{eq:Cycle01} by:
\begin{equation}
x_c+ \sum_{s\in\bigcup_{i\in V(c)} S(i,c)} x_s\geq 1, \forall c\in \setC, \label{eq:CyclePrefStrict}
\end{equation}
Constraints~\eqref{eq:CyclePrefStrict} guarantee that either $c$ is in the exchange or one of its vertices is matched in a cycle strictly better than $c$.

Finally, for the case of weak preferences, let us define $E_{(i,c)}=B_{(i,c)}\setminus S_{(i,c)}$, that is, $E_{(i,c)}=\{\bar c \in \setC(i), \bar c\neq c : \bar c \sim_i c\}$ is a set of cycles that are different from $c$ and are equally good for $i$ (see Definition~\ref{def_indif_cyc}).   Constraints~(\ref{eq:CyclePrefStrict}) can be modified as follows:
\begin{equation}
x_c+\frac{1}{|c|}\sum_{i \in V(c)} \sum_{s\in E_{(i,c)}} x_s  + \sum_{s\in\bigcup_{i\in V(c)} S(i,c)} x_s\geq 1, \forall c\in \setC, \label{eq:s_stab_cycle_ties}
\end{equation}
If for a cycle $c$ there is a vertex $i\in V(c)$ such that there is no cycle $s$ different form $c$ where $s\sim_i c$, then the latter constraint can be tightened for that cycle in the following way:
\begin{equation}
x_c  + \sum_{s\in\bigcup_{i\in V(c)} S(i,c)} x_s\geq 1,   \label{eq:s_stab_cycle_ties_tight}
\end{equation}



\par
\subsection{Cycle-Edge Formulation}
\par

Preliminary experiments identified some weaknesses on the two previous formulations. The bottleneck for the edge formulation is the large number of path constraints~\eqref{eq:EdgePathC} and~\eqref{eq:EdgePathN}. For the cycle formulation the stability constraints are large in size as they contain one variable for each cycle that shares the same arc $(i,j)$. For the same purpose, variable $y_{ij}$  in the edge formulation only  appears in the constraint when necessary.

We now propose an alternative formulation, called cycle-edge formulation, where we retain advantages from both formulations. This new formulation uses both integer variables $x$ and $y$ in a consistent way: for every cycle $c\in\setC$, we require that $x_c=1$ if and only if $y_{ij}=1$ for every $(i,j)\in A(c)$. This correspondence can be achieved by the basic feasibility cycle-constraints \eqref{eq:CycleMostOne},
and by adding the following set of constraints:
\begin{equation}
    \sum_{c: (i,j)\in A(c)} x_c = y_{ij}, \forall (i,j)\in A \label{eq:cycle-edge_arceq}
\end{equation}

Bearing in mind the packing constraints~\eqref{eq:CycleMostOne}, we may relax the binary requirement for variables $y_{ij}$, and consider:
\begin{equation}
    y_{ij} \geq 0\quad \forall (i,j)\in A.
\end{equation}

Stability and strong stability are assured by constraints \eqref{eq:stab_edge} and \eqref{eq:s_stab_edge} or \eqref{eq:s_stab_edge_ties}, respectively.
The objective function can be represented either by  (\ref{eq:objcycle}) or (\ref{eq:EdgeObj}).

\section{Relaxing stability}\label{sec:nearly_stable}


Since finding a stable exchange in a graph is not always possible, in this section we discuss the possibility of obtaining ``least non-stable" solutions when stability is not achievable. The aim in such cases will be to find the exchange with minimum number of blocking cycles that  has maximum cardinality. Furthermore, since stability may negatively impact the maximum number of transplants, we also discuss the possibility of relaxing stability constraints so that the decrease in the number of transplants does not go over a pre-defined limit.



Let $a_c u \geq b_c$  $\forall c\in \setC$, be one of the sets of constraints
associated to stability in Section \ref{sec:ip} (constraints (\ref{eq:stab_edge}), (\ref{eq:s_stab_edge}), (\ref{eq:s_stab_edge_ties}), (\ref{eq:CyclePref}), (\ref{eq:CyclePrefStrict}), or (\ref{eq:s_stab_cycle_ties})), where $u$ is a vector of corresponding variables, $x$ or $y$.
Consider also the slack variables  $\tau_c\in\{0,1\}$ that will indicate whether the stability constraint associated to cycle $c$ is violated, or not (i.e. whether cycle $c$ is blocking). 

The following IP model is to be solved:

\begin{align}
& & \min_{u,\tau}&&& |V| \sum_{c\in\mathcal{C}}\, \tau_c - \omega u \label{eq:objunstable}\\
& & 	&&& a_c u + b_c\tau_c \geq b_c 			&&  \forall c\in \setC \label{eq:BlockCycle} \\
&& &&& \mbox{(\emph{Other model constraints})}  \label{eq:BlockCycleOthers}
\end{align}
where \emph{(Other model constraints)} refers to all constraints, except for the stability ones, in the edge, cycle and cycle-edge formulations.
Coefficients $\omega$ in the objective function~\eqref{eq:objunstable} correspond to coefficients of the respective objective function: arc weight and cycle weight, for the edge and cycle formulations, respectively.
Coefficient $|V|$ in the first term of  (\ref{eq:objunstable}) guarantees the minimization of the number of blocking cycles in the solution at first place. Indeed, for any solution $u$ the value of the second term $\omega u$ will never exceed the number of vertices. Constraints (\ref{eq:BlockCycle}) signal if a blocking cycle is in the solution. 






In the context of a KEP, requesting for solution stability may represent a decrease in the maximum number of transplants in a pool. In such a case, if the decrease goes over a certain value, it may happen that the decision maker would wish to relax the stability of the solution  to keep the number of transplants within what he/she considers to be an acceptable distance from the maximum cardinality solution. This can be addressed by  solving the problem (\ref{eq:objunstable})--(\ref{eq:BlockCycleOthers}), with the following additional constraint:

\begin{equation}
\sum_{c:c\in\mathcal{C}}\,|c|\cdot x_c  \geq M^* - R \qquad (\sum_{(i,j)\in A} w_{ij} y_{ij} \geq M^* - R) \label{eq:constr_loose}\\
\end{equation}

\noindent where $M^*$ is the maximum number of transplants that can be achieved if stability is not imposed, and $R$ is the maximum number of transplants we accept to loose in order to have a ``more stable" solution.
Considering different values of $R$ the decision maker may study different trade-offs between the two objectives.
Notice that when $R=M^*$ the problem corresponds to finding a stable solution or, for the cases where a stable solution does not exist, to the minimization of the number of blocking cycles. On the other hand, by solving the problem~\eqref{eq:objunstable}--\eqref{eq:constr_loose} with $R=0$ one will find an exchange with maximum number of transplants that has minimal number of blocking cycles.
\par

\section{Computational experiments}\label{sec:comp}

In this section we present computational results for validation and comparison of the models proposed in Sections~\ref{sec:ip} and \ref{sec:nearly_stable}.
To refer to the formulations we use CF for the cycle formulation, EF for the edge formulation, and
CEF for the cycle-edge formulation.

All formulations were implemented using Python programming language and tested using Gurobi as optimization solver~\cite{gurobi}. The tests were executed on a MacMini 8 running macOS version
10.14.3 in a Intel Core i7 CPU with 6 cores at 3.2 GHz with 8GB of RAM.

\subsection{Test instances}
 In our analysis we use test instances generated with the generator proposed in~\cite{Santos17,fair19}. Each instance mimics a pool of a KEP: patients and donors are generated with their characteristics (blood-type and sensitization level) and added to the pool if the generated pair is considered incompatible. An instance has a given number incompatible pairs ($P$) and the number of non-directed donors is $|N| = \lceil 5\% |P| \rceil$. Fifty instances of each size were considered.
Preferences were generated randomly. For weak preferences, random weights in the interval $(0,1)$ were assigned to each arc. Incoming arcs with weights within each interval of length 0.1 were considered equally preferable. We imposed the maximum length of cycles and chains to be equal, i.e. $K=L$, and performed experiments for $K= 2,3$ and 4.

Table~\ref{tab:inst} summarizes the characteristics of the instances considered: number of NDDs $|N|$; average number of arcs in instances of a given size  $|A|$ ; average number of cycles $|\mathcal{C}^P|_K$ and  of chains $|\mathcal{C}^N|_K$ for each $K$; and CPU time $t_{K}$ required for enumeration of cycles and chains.
\begin{sidewaystable}[]
    \centering
    \tiny
    \begin{tabular}{c|rrrrrrrrrrrrrrrrrr}
    \cline{1-19}
$|P|$        	&	20	&	30	&	40	&	50	&	60	&	70	&	80	&	90	&	100	&	150	&	160	&	170	&	180	&	200	&	250	&	300	&	350	&	400	\\
$|N|$	&	1	&	2	&	2	&	3	&	3	&	4	&	4	&	5	&	5	&	8	&	8	&	9	&	9	&	10	&	13	&	15	&	18	&	20	\\
$|A|$	&	98	&	225	&	392	&	632	&	901	&	1242	&	1610	&	2070	&	2520	&	5738	&	6491	&	7362	&	8244	&	10104	&	15941	&	23054	&	31397	&	41019	\\
\cline{1-19}
$|\mathcal{C}^P|_2$	&	9	&	19	&	36	&	56	&	79	&	108	&	144	&	187	&	232	&	538	&	612	&	694	&	783	&	952	&	1487	&	2196	&	2983	&	3928	\\
$|\mathcal{C}^N|_2$	&	8	&	23	&	30	&	59	&	70	&	109	&	124	&	177	&	197	&	458	&	487	&	578	&	612	&	752	&	1231	&	1728	&	2378	&	3020	\\
$|\mathcal{C}^P|_3$	&	27	&	82	&	192	&	376	&	599	&	934	&	1416	&	2074	&	2824	&	9813	&	11881	&	14335	&	17067	&		&		&		&		&		\\
$|\mathcal{C}^N|_3$	&	37	&	151	&	260	&	637	&	885	&	1587	&	2079	&	3335	&	4071	&	14156	&	16003	&	20202	&	22646	&		&		&		&		&		\\
$|\mathcal{C}^P|_4$	&	72	&	339	&	1136	&	2843	&	5337	&	9752	&	17139	&	28530	&	43553	&		&		&		&		&		&		&		&		&		\\
$|\mathcal{C}^N|_4$	&	147	&	874	&	2062	&	6242	&	10240	&	21364	&	32475	&	60088	&	81835	&		&		&		&		&		&		&		&		&		\\
\cline{1-19}
$t_2$	&	0.0	&	0.0	&	0.0	&	0.0	&	0.0	&	0.0	&	0.0	&	0.0	&	0.0	&	0.1	&	0.1	&	0.1	&	0.1	&	0.2	&	0.5	&	1.0	&	1.9	&	3.2	\\
$t_3$	&	0.0	&	0.0	&	0.0	&	0.0	&	0.0	&	0.1	&	0.2	&	0.4	&	0.6	&	6.6	&	8.8	&	13.3	&	17.5	&		&		&		&		&		\\
$t_4$	&	0	&	0	&	0.2	&	1.2	&	3.3	&	12.5	&	30.9	&	97.5	&	194.4	&		&		&		&		&		&		&		&		&		\\
\cline{1-19}
    \end{tabular}
  \caption{Characteristics of instances for different values of maximum length of cycles $K$. }
    \label{tab:inst}
\end{sidewaystable}

As expected the number of cycles and chains increases exponentially when $K$ increases. We can also observe that a main difficulty of the instances is associated to the  presence of NDDs: there are almost twice as many chains as cycles for $K=3,4$. Nevertheless the CPU time required for enumeration of cycles and chains is reasonable even for larger instances.

\subsection{Comparison of formulations}\label{sec:comp_formulations}

The formulations proposed in Section~\ref{sec:ip} were tested on different instances, for different values of $K$. For the case of the cycle-edge formulation, results are presented using both objective functions~\eqref{eq:objcycle} and \eqref{eq:EdgeObj}.

Tables~\ref{tab:strict} and~\ref{tab:weak} report average results for each set of instances of the same size,  for  strict (Table~\ref{tab:strict}) and weak (Table~\ref{tab:weak}) preferences, respectively. The following notation is used in the tables:

\begin{itemize}
    \item[--] $G$ is the average gap of the linear relaxation of a formulation for instances of a given size;
    \item[--] $T$ is the average CPU time (in seconds) needed to find an optimal solution; maximum CPU time of 1 hour was imposed; the number of instances that were not solved within the limit is presented in between parenthesis; for the case of the CEF formulation we denote by $T_x$ the average CPU time  when the objective function~\eqref{eq:objcycle} (with $x$ variables) is used and by $T_y$ the average CPU time for objective~\eqref{eq:EdgeObj} (with $y$ variables).
    \item[--] $\#I$ is the number of instances out of 50 that do not have a stable or a strongly stable solution.
    \item [--]  $ps$ is the average \emph{price of stability}, i.e. the  relative loss in number of transplants when compared to the maximum value: $ps = \frac{M^* - M^s}{M^*} \dot 100 \%$, where $M^*$ is the maximum number of transplants for a given instance and $M^s$ is the number of transplants in a maximum cardinality stable or strongly stable solution (calculated only for instances where a solution exists).
\end{itemize}

\begin{sidewaystable}

\tiny
    \centering
\begin{tabular}{l|rr|rr|rr|rrr||rr|rr|rr|rrr}
&\multicolumn{9}{c||}{Stable} & \multicolumn{9}{c}{Strongly Stable}\\
\cline{2-19}
& & &\multicolumn{2}{c|}{CF} & \multicolumn{2}{c|}{EF} & \multicolumn{3}{c||}{CEF}
& & &\multicolumn{2}{c|}{CF} & \multicolumn{2}{c|}{EF} & \multicolumn{3}{c}{CEF}\\
$|P|$	&	
$\#I$ & $ps.$ &
$G$	 &   $T$	& $G$	 &   $T$	& $G$	  &   $T_x$	&  $T_y$ &
$\#I$ & $ps.$ &
$G$	 &   $T$	& $G$	 &   $T$	& $G$	  &   $T_x$	&  $T_y$ \\
\cline{1-19}
\multicolumn{19}{c}{\pmb{$K=2$}}\\
20	&	0	&	4.6	&	0	&		0.0	&	2.9	&		0.0	&	2.8	&		0.0	&		0.0	&	0	&	4.6	&	0	&		0.0	&	0.3	&		0.0	&	0	&		0.0	&		0.0	\\
30	&	1	&	5.7	&	0	&		0.0	&	3.8	&		0.0	&	3.6	&		0.0	&		0.0	&	1	&	5.7	&	0	&		0.0	&	0	&		0.0	&	0	&		0.0	&		0.0	\\
40	&	1	&	5.7	&	0	&		0.0	&	3.8	&		0.0	&	3.7	&		0.0	&		0.0	&	1	&	5.7	&	0	&		0.0	&	0.1	&		0.0	&	0	&		0.0	&		0.0	\\
50	&	2	&	5.6	&	0	&		0.0	&	3.6	&		0.1	&	3.4	&		0.0	&		0.0	&	2	&	5.6	&	0	&		0.0	&	0	&		0.0	&	0	&		0.0	&		0.0	\\
60	&	2	&	7.2	&	0	&		0.0	&	4.7	&		0.1	&	4.4	&		0.0	&		0.0	&	2	&	7.2	&	0	&		0.0	&	0.2	&		0.0	&	0	&		0.0	&		0.0	\\
70	&	3	&	7.6	&	0	&		0.0	&	5.3	&		0.3	&	4.9	&		0.0	&		0.0	&	3	&	7.6	&	0	&		0.0	&	0.3	&		0.1	&	0	&		0.0	&		0.0	\\
80	&	3	&	8.0	&	0	&		0.0	&	5.8	&		0.5	&	5.2	&		0.0	&		0.0	&	3	&	8.0	&	0	&		0.0	&	0.8	&		0.2	&	0	&		0.0	&		0.0	\\
90	&	6	&	9.3	&	0	&		0.0	&	6.8	&		1.0	&	6.2	&		0.0	&		0.0	&	6	&	9.3	&	0	&		0.0	&	1.1	&		0.4	&	0	&		0.0	&		0.0	\\
100	&	1	&	10.0	&	0	&		0.0	&	7.3	&		1.5	&	6.7	&		0.0	&		0.0	&	1	&	10.0	&	0	&		0.0	&	1.9	&		0.6	&	0	&		0.0	&		0.0	\\
150	&	2	&	9.8	&	0	&		0.0	&	7.4	&		10.3	&	6.3	&		0.0	&		0.0	&	2	&	9.8	&	0	&		0.0	&	2.7	&		6.0	&	0	&		0.1	&		0.1	\\
160	&	3	&	9.5	&	0	&		0.1	&	7.2	&		13.6	&	6.1	&		0.0	&		0.0	&	3	&	9.5	&	0	&		0.1	&	2.9	&		8.7	&	0	&		0.1	&		0.1	\\
170	&	8	&	9.9	&	0	&		0.1	&	7.2	&		18.9	&	6	&		0.1	&		0.0	&	8	&	9.9	&	0	&		0.1	&	3.3	&		13.4	&	0	&		0.1	&		0.1	\\
180	&	3	&	10.0	&	0	&		0.1	&	7.5	&		26.9	&	6.4	&		0.1	&		0.1	&	3	&	10.0	&	0	&		0.1	&	3.5	&		17.0	&	0	&		0.1	&		0.1	\\
200	&	5	&	10.1	&	0	&		0.1	&	7.7	&		42.2	&	6.5	&		0.1	&		0.1	&	5	&	10.1	&	0	&		0.1	&	4.1	&		29.2	&	0	&		0.2	&		0.2	\\
250	&	8	&	9.5	&	0	&		0.2	&		&			&	5.9	&		0.2	&		0.2	&	8	&	9.5	&	0	&		0.2	&		&			&	0	&		0.3	&		0.3	\\
300	&	4	&	10.0	&	0	&		0.3	&		&			&	6.5	&		0.3	&		0.3	&	4	&	10.0	&	0	&		0.3	&		&			&	0	&		0.5	&		0.5	\\
350	&	11	&	10.0	&	0	&		0.5	&		&			&	6.4	&		0.6	&		0.6	&	11	&	10.0	&	0	&		0.6	&		&			&	0	&		1.0	&		0.8	\\
400	&	18	&	10.0	&	0	&		0.8	&		&			&	6.3	&		1.0	&		1.0	&	18	&	10.0	&	0	&		0.8	&		&			&	0	&		1.4	&		1.2	\\

\cline{1-19}			
\multicolumn{19}{c}{\pmb{$K=3$}}\\
20	&		&	9.0	&	0.2	&		0.0	&	6.9	&		0.0	&	3.5	&		0.0	&		0.0	&	2	&	18.3	&	0	&		0.0	&	4.2	&		0.0	&	0.2	&		0.0	&		0.0	\\
30	&		&	11.5	&	0.5	&		0.0	&	11.2	&		0.7	&	5.5	&		0.0	&		0.0	&	1	&	20.4	&	0	&		0.0	&	8.4	&		0.1	&	0.2	&		0.0	&		0.0	\\
40	&		&	10.0	&	0.3	&		0.1	&	9.2	&		4.1	&	4.5	&		0.0	&		0.0	&	3	&	18.4	&	0	&		0.0	&	10.4	&		0.2	&	0.3	&		0.0	&		0.0	\\
50	&		&	10.9	&	0.5	&		0.2	&	11.1	&		41.1	&	5.8	&		0.1	&		0.1	&	4	&	21.3	&	0	&		0.1	&	14.5	&		1.0	&	0.9	&		0.1	&		0.1	\\
60	&		&	10.9	&	0.4	&		0.4	&	10	&		192.8	&	5	&		0.2	&		0.2	&	4	&	21.4	&	0.3	&		0.3	&	15.2	&		2.5	&	0.8	&		0.1	&		0.2	\\
70	&		&	11.6	&	0.8	&		1.4	&	11.1	&	(4)	651.5	&	5.7	&		0.5	&		0.4	&	3	&	20.8	&	0.1	&		0.8	&	17.1	&		17.6	&	0.6	&		0.2	&		0.4	\\
80	&		&	10.8	&	0.9	&		2.3	&		&			&	5.9	&		1.0	&		1.0	&	9	&	20.3	&	0.1	&		1.5	&	19.2	&		49.4	&	1.2	&		0.5	&		0.8	\\
90	&		&	9.4	&	1.2	&		6.7	&		&			&	5.4	&		2.9	&		3.0	&	5	&	18.6	&	0.1	&		2.9	&	17.3	&	(1)	198.1	&	1.7	&		1.2	&		1.8	\\
100	&		&	9.2	&	1.2	&		11.1	&		&			&	5.2	&		5.1	&		4.8	&	4	&	18.1	&	0	&		4.4	&	18	&	(3)	366.2	&	2.2	&		1.7	&		2.7	\\
150	&		&	6.8	&	1.5	&	(1)	294.9	&		&			&	4.4	&		165.7	&		202.4	&	10	&	15.5	&	0.5	&		53.9	&		&			&	5	&		33.9	&		52.5	\\
160	&		&	7.2	&	1.7	&	(2)	521.2	&		&			&	4.8	&	(1)	339.3	&	(2)	308.2	&	10	&	17.1	&	1	&		75.4	&		&			&	6	&		55.0	&		82.5	\\
170	&		&	7.0	&	1.9	&	(6)	872.8	&		&			&	4.6	&	(2)	778.6	&	(2)	751.5	&	9	&	15.7	&	1.4	&		171.4	&		&			&	5.7	&		106.0	&		143.1	\\
180	&		&	7.2	&	1.9	&	(11)	963.6	&		&			&	4.7	&	(4)	913.4	&	(7)	842.7	&	15	&	16.0	&	0.9	&		221.1	&		&			&	5.8	&		145.3	&		208.4	\\
\cline{1-19}	
\multicolumn{19}{c}{\pmb{$K=4$}}
\\20	&		&	8.7	&	0.6	&		0.0	&		&			&	3.1	&		0.0	&		0.0	&	0	&	22.1	&	0	&		0.0	&		&			&	0.2	&		0.0	&		0.0	\\
30	&		&	8.7	&	0.6	&		1.0	&		&			&	4.1	&		0.1	&		0.1	&	2	&	23.0	&	0.6	&		0.5	&		&			&	1.6	&		0.1	&		0.1	\\
40	&		&	8.0	&	0.8	&		8.5	&		&			&	3.7	&		0.7	&		0.6	&	2	&	20.5	&	0	&		3.3	&		&			&	1.8	&		0.4	&		0.5	\\
50	&		&	7.2	&	0.9	&		78.5	&		&			&	3.9	&		3.9	&		4.0	&	2	&	21.6	&	0.4	&		23.6	&		&			&	5.2	&		2.4	&		3.0	\\
60	&		&	7.4	&	1.2	&	(1)	206.7	&		&			&	4.1	&		15.0	&		16.7	&	4	&	22.4	&	0.2	&		92.3	&		&			&	5.9	&		8.0	&		10.3	\\
    \end{tabular}
  \caption{Comparison of formulations for instances with strict preferences.}
    \label{tab:strict}
\end{sidewaystable}

\begin{sidewaystable}

\tiny
    \centering
    \hspace*{-0.5cm}
\begin{tabular}{l|rr|rr|rr|rrr||rr|rr|rr|rrr}
&\multicolumn{9}{c||}{Stable} & \multicolumn{9}{c}{Strongly Stable}\\
\cline{2-19}
& & &\multicolumn{2}{c|}{CF} & \multicolumn{2}{c|}{EF} & \multicolumn{3}{c||}{CEF}
& & &\multicolumn{2}{c|}{CF} & \multicolumn{2}{c|}{EF} & \multicolumn{3}{c}{CEF}\\
$|P|$	&	
$\#I$ & $ps.$ &
$G$	 &   $T$	& $G$	 &   $T$	& $G$	  &   $T_x$	&  $T_y$ &
$\#I$ & $ps.$ &
$G$	 &   $T$	& $G$	 &   $T$	& $G$	  &   $T_x$	&  $T_y$ \\
\cline{1-19}
\multicolumn{19}{c}{\pmb{$K=2$}}\\

20	&	0	&	3.8	&	0	&		0.0	&	2.43	&		0.0	&	2.41	&		0.0	&		0.0	&	11	&	3.9	&	0	&		0.0	&	0	&		0.0	&	0	&	0.0	&	0.0	\\
30	&	1	&	5.3	&	0	&		0.0	&	3.59	&		0.0	&	3.41	&		0.0	&		0.0	&	17	&	4.2	&	0	&		0.0	&	0	&		0.0	&	0	&	0.0	&	0.0	\\
40	&	1	&	5.2	&	0	&		0.0	&	3.56	&		0.0	&	3.44	&		0.0	&		0.0	&	19	&	5.9	&	0	&		0.0	&	0.04	&		0.0	&	0	&	0.0	&	0.0	\\
50	&	1	&	5.3	&	0.09	&		0.0	&	3.38	&		0.1	&	3.24	&		0.0	&		0.0	&	28	&	5.2	&	0	&		0.0	&	0	&		0.0	&	0	&	0.0	&	0.0	\\
60	&	1	&	6.9	&	0.05	&		0.0	&	4.55	&		0.1	&	4.21	&		0.0	&		0.0	&	23	&	7.6	&	0	&		0.0	&	0.13	&		0.0	&	0	&	0.0	&	0.0	\\
70	&	3	&	7.0	&	0	&		0.0	&	4.81	&		0.3	&	4.48	&		0.0	&		0.0	&	27	&	6.4	&	0	&		0.0	&	0.59	&		0.1	&	0	&	0.0	&	0.0	\\
80	&	3	&	7.7	&	0	&		0.0	&	5.64	&		0.5	&	5.04	&		0.0	&		0.0	&	25	&	7.8	&	0	&		0.0	&	0.6	&		0.2	&	0	&	0.0	&	0.0	\\
90	&	6	&	8.9	&	0.09	&		0.0	&	6.48	&		1.0	&	5.92	&		0.0	&		0.0	&	27	&	10.7	&	0	&		0.0	&	0.75	&		0.3	&	0	&	0.0	&	0.0	\\
100	&	0	&	9.4	&	0.03	&		0.0	&	6.89	&		1.6	&	6.34	&		0.0	&		0.0	&	34	&	9.9	&	0	&		0.0	&	2.47	&		0.7	&	0	&	0.0	&	0.0	\\
150	&	2	&	9.4	&	0.03	&		0.0	&	7.02	&		10.5	&	5.99	&		0.0	&		0.0	&	37	&	9.6	&	0	&		0.0	&	1.18	&		4.8	&	0	&	0.1	&	0.1	\\
160	&	2	&	9.3	&	0	&		0.1	&	6.94	&		14.0	&	5.91	&		0.0	&		0.0	&	42	&	7.4	&	0	&		0.1	&	2.36	&		7.7	&	0	&	0.1	&	0.1	\\
170	&	8	&	9.7	&	0.03	&		0.1	&	7.09	&		19.8	&	5.93	&		0.1	&		0.0	&	40	&	8.0	&	0	&		0.1	&	2.37	&		11.6	&	0	&	0.1	&	0.1	\\
180	&	3	&	9.7	&	0	&		0.1	&	7.16	&		27.1	&	6.11	&		0.1	&		0.1	&	34	&	9.5	&	0	&		0.1	&	2.8	&		14.2	&	0	&	0.1	&	0.1	\\
200	&	3	&	9.8	&	0.06	&		0.1	&	7.51	&		45.2	&	6.36	&		0.1	&		0.1	&	38	&	11.1	&	0	&		0.1	&	4.45	&		25.6	&	0	&	0.2	&	0.2	\\
250	&	8	&	9.3	&	0	&		0.2	&		&			&	5.76	&		0.2	&		0.2	&	46	&	9.6	&	0	&		0.2	&		&			&	0	&	0.3	&	0.3	\\
300	&	2	&	10.0	&	0.01	&		0.3	&		&			&	6.41	&		0.3	&		0.3	&	43	&	11.0	&	0	&		0.3	&		&			&	0	&	0.5	&	0.5	\\
350	&	10	&	9.9	&	0.03	&		0.5	&		&			&	6.29	&		0.6	&		0.6	&	48	&	10.2	&	0	&		0.6	&		&			&	0	&	1.0	&	0.8	\\
400	&	15	&	10.0	&	0.03	&		0.8	&		&			&	6.31	&		1.1	&		1.0	&	50	&	-	&		-&		0.8	&	-	&			&	-	&	1.4	&	1.2	\\

\cline{1-19}
\multicolumn{19}{c}{\pmb{$K=3$}}\\
20	&		&	8.5	&	0.23	&		0.0	&	6.82	&		0.0	&	3.64	&		0.0	&		0.0	&	18	&	20.6	&	0.3	&		0.0	&	2.95	&		0.0	&	0.3	&	0.0	&	0.0	\\
30	&		&	10.5	&	0.5	&		0.0	&	10.12	&		1.6	&	4.71	&		0.0	&		0.0	&	30	&	21.2	&	0	&		0.0	&	5.68	&		0.1	&	0	&	0.0	&	0.0	\\
40	&		&	9.6	&	0.6	&		0.1	&	8.89	&		4.5	&	4.3	&		0.0	&		0.0	&	25	&	18.2	&	0.32	&		0.1	&	10.09	&		0.2	&	0.32	&	0.0	&	0.0	\\
50	&		&	9.6	&	0.55	&		0.2	&	9.6	&		40.9	&	4.83	&		0.1	&		0.1	&	32	&	21.3	&	0.39	&		0.2	&	13.02	&		1.0	&	0.39	&	0.1	&	0.1	\\
60	&		&	10.3	&	0.54	&		0.5	&	9.51	&		202.5	&	4.76	&		0.2	&		0.2	&	29	&	22.3	&	0.56	&		0.4	&	14.66	&		5.4	&	0.56	&	0.1	&	0.2	\\
70	&		&	11.0	&	0.82	&		1.4	&	10.44	&	(4)	686.0	&	5.45	&		0.6	&		0.6	&	33	&	19.8	&	0.41	&		1.0	&	14.82	&		20.9	&	0.41	&	0.2	&	0.4	\\
80	&		&	10.1	&	0.74	&		2.3	&		&			&	5.44	&		1.1	&		1.2	&	37	&	19.4	&	0.07	&		1.7	&	17.9	&		24.0	&	0.07	&	0.4	&	0.7	\\
90	&		&	8.6	&	1.04	&		6.9	&		&			&	4.77	&		3.8	&		4.0	&	41	&	14.4	&	0.65	&		4.4	&	12.17	&		111.6	&	0.66	&	1.0	&	1.5	\\
100	&		&	8.6	&	1.16	&		13.8	&		&			&	4.84	&		5.8	&		6.2	&	41	&	20.8	&	2.42	&		5.7	&	19.88	&	(2)	240.1	&	2.42	&	1.5	&	2.5	\\
150	&		&	6.4	&	1.6	&	(1)	377.1	&		&			&	4.14	&		258.5	&		286.2	&	46	&	16.4	&	5.03	&		86.5	&		&			&	5.03	&	36.3	&	76.6	\\
160	&		&	6.6	&	1.61	&		680.1	&		&			&	4.27	&	(1)	389.1	&	(2)	360.4	&	43	&	15.0	&	5.32	&		119.8	&		&			&	5.32	&	60.7	&	122.3	\\
170	&		&	6.4	&	1.63	&	(4)	985.9	&		&			&	4.08	&	(5)	594.1	&	(4)	741.6	&	47	&	10.7	&	2.95	&		397.7	&		&			&	2.95	&	113.0	&	301.3	\\
180	&		&	6.6	&	1.69	&	(8)	1103.2	&		&			&	4.25	&	(5)	833.4	&	(8)	790.1	&	47	&	11.0	&	2.71	&	(2)	424.7	&		&			&	2.71	&	203.4	&	362.6	\\

\cline{1-19}
\multicolumn{19}{c}{\pmb{$K=4$}}\\
20	&		&	8.1	&	0.64	&		0.0	&		&			&	2.81	&		0.0	&		0.0	&	17	&	24.2	&	0	&		0.0	&		&			&	0	&	0.0	&	0.0	\\	
30	&		&	8.1	&	0.56	&		0.8	&		&			&	3.52	&		0.1	&		0.2	&	31	&	23.8	&	0.38	&		1.1	&		&			&	0.38	&	0.1	&	0.1	\\	
40	&		&	7.2	&	0.79	&		9.5	&		&			&	3.27	&		0.7	&		0.7	&	28	&	18.9	&	0.97	&		5.9	&		&			&	0.97	&	0.2	&	0.4	\\	
50	&		&	6.9	&	1.04	&		79.7	&		&			&	3.75	&		4.2	&		4.0	&	34	&	20.9	&	2.64	&		62.1	&		&			&	2.65	&	2.1	&	3.3	\\	
60	&		&	6.7	&	1.2	&	(2)	218.1	&		&			&	3.56	&		18.8	&		18.0	&	34	&	22.6	&	5.54	&		240.8	&		&			&	5.54	&	7.3	&	10.0	\\

    \end{tabular}
  \caption{Comparison of formulations for instances with weak preferences.}
    \label{tab:weak}
\end{sidewaystable}

Results in Table~\ref{tab:strict} show that the EF is not competitive when compared to the other formulations,  both for stability and strong stability. Tests were run only for instances of up to 200 vertices for $K=2$, and up to 60 and 100 vertices for $K=3$. For instances of those sizes, $T$ for the EF is already drastically larger than the one obtained for the other formulations. For $K=4$ the EF was not run.

In terms of gap of the linear relaxation (columns $G$) the CF has the best performance, with average values of at most 2\% for all instances and values of $K$. For  $K=2$, a stable solutions is always strongly stable. The CF is tight even in case of stability, having zero gap for all instances, while the stable CEF formulation is less tight than the strongly stable CEF (compare columns $G$ for CEF). 
Notice that the EF is the least tight among all the formulations, having the largest gap of the linear relaxation. This observation, together with the large number of constraints~\eqref{eq:EdgePathC} associated to this formulation, justifies the higher CPU times.

Despite of having a larger gap than for CF, CEF is computationally more efficient than CF for the two objective functions considered. For $K=3,4$, and larger instances ($|P|\geq 150$ and  $|P|=60$, respectively) CEF found stable solutions for more instances  within the time limit for the stable exchange problem. Also CEF was faster for  instances of smaller size and for the strongly stable problem. The difference between CEF and CF becomes even more evident when $K=4$.
Experiments on instances of size bigger than 60 were not run due to high memory consumption by the solver.

In what regards the existence of a solution (see columns $\#I$), for the case of stability and $K=3,4$ there were no instances without a stable solution. Differently, for strong stability, there was a considerable number of instances with no solution. Moreover, to obtain a strongly stable solution one may have to sacrifice more than 20\% of the transplants, while for stable solution the price of stability is less than 11\% (compare columns $ps.$).
Finally, we may observe that finding strongly stable exchanges is slightly easier than finding stable exchanges:  all formulations need shorter CPU times and more instances are solved within the same time limit (compare the performance of each formulation in columns ``Stable" and ``Strongly Stable" for $K = 3$, $|P| > 100$, $K=4$, $|P|>40$).

Comparing the results for weak preferences presented in Table~\ref{tab:weak} with those in Table~\ref{tab:strict} we reach similar conclusions. The performance of all the models is in general very similar, when compared to the same instances with strict preferences. The main difference between the tables is in the number of instances where a solution exists (columns $I$). In particular, for the case of weak preferences,  there are significantly more instances where a strongly stable solution does not exist, e.g.  no instances with 400 pairs and $K=2$ had a strongly stable solution. 

Interestingly, for both strict and weak preferences we may observe a significant difference in the performance of the CEF for the two alternative objective functions: CEF with $x$-objective is the fastest among all the formulations at finding a strongly stable exchange (see column $T_x$ in "Strongly Stable" columns of Tables~\ref{tab:strict} and~\ref{tab:weak}).

{\subsection{Analyzing the trade-off \emph{stability versus number of transplants}}}

Following the discussion in Section~\ref{sec:nearly_stable}, in this section we will analyze how stability requirements impact the objective of maximizing the number of transplants.  
For doing so, we will compare two extreme solutions in terms of the number of transplants achieved: the solution that maximizes the number of transplants and the stable solution (or solution with minimum number of blocking cycles when a stable solution does not exist). Furthermore, we  will evaluate intermediate cases that represent different trade-offs between achieving stability and maximizing the number of transplants. This is achieved by running the models presented in Section~\ref{sec:nearly_stable}, setting  $R$ in constraints~\eqref{eq:constr_loose} to $\kappa M^*$, for different values of $\kappa$, and rounding up the right hand side of the same constraints in order to get an integer value: $\lceil M^* - \kappa M^*\rceil$.

Based on the results in Tables~\ref{tab:strict} and~\ref{tab:weak}, where the average value of the price of stability ($ps.$) was at most 24\%, we considered the following set of values for parameter $\kappa : \{0.05, 0.1, 0.15, 0.2 \}$; for $\kappa \geq$ 0.25 we will, in general, get stable solutions. Note that for $\kappa = 0$ we obtain the solution with maximum number of transplant that has minimum number of blocking cycles, and for $\kappa = 1$ we obtain a stable solution (or a solution with minimum number of blocking cycles, if there is no stable solution)  with maximum number of transplants. 

\begin{figure}[htbp]
\centering
\includegraphics[scale = 0.6]{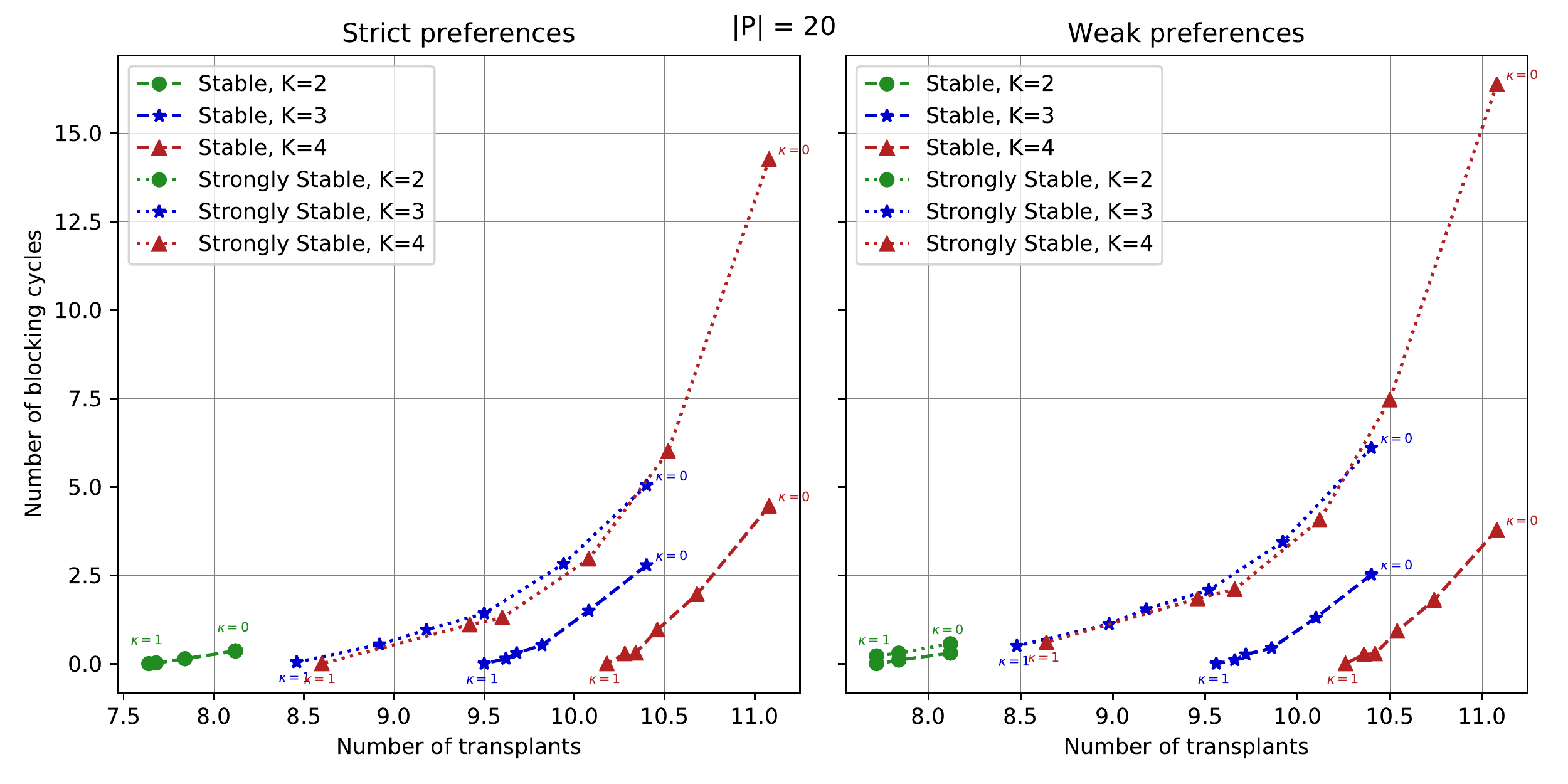}
\caption{Trade-off between stability  and  maximum  number of transplants for instances with $|P| = 20$; 
$\kappa = 0, 0.05,0.1, 0.15, 0.2, 1$.}\label{fig:20ps}
\end{figure}

\begin{figure}[htbp]
\includegraphics[scale = 0.6]{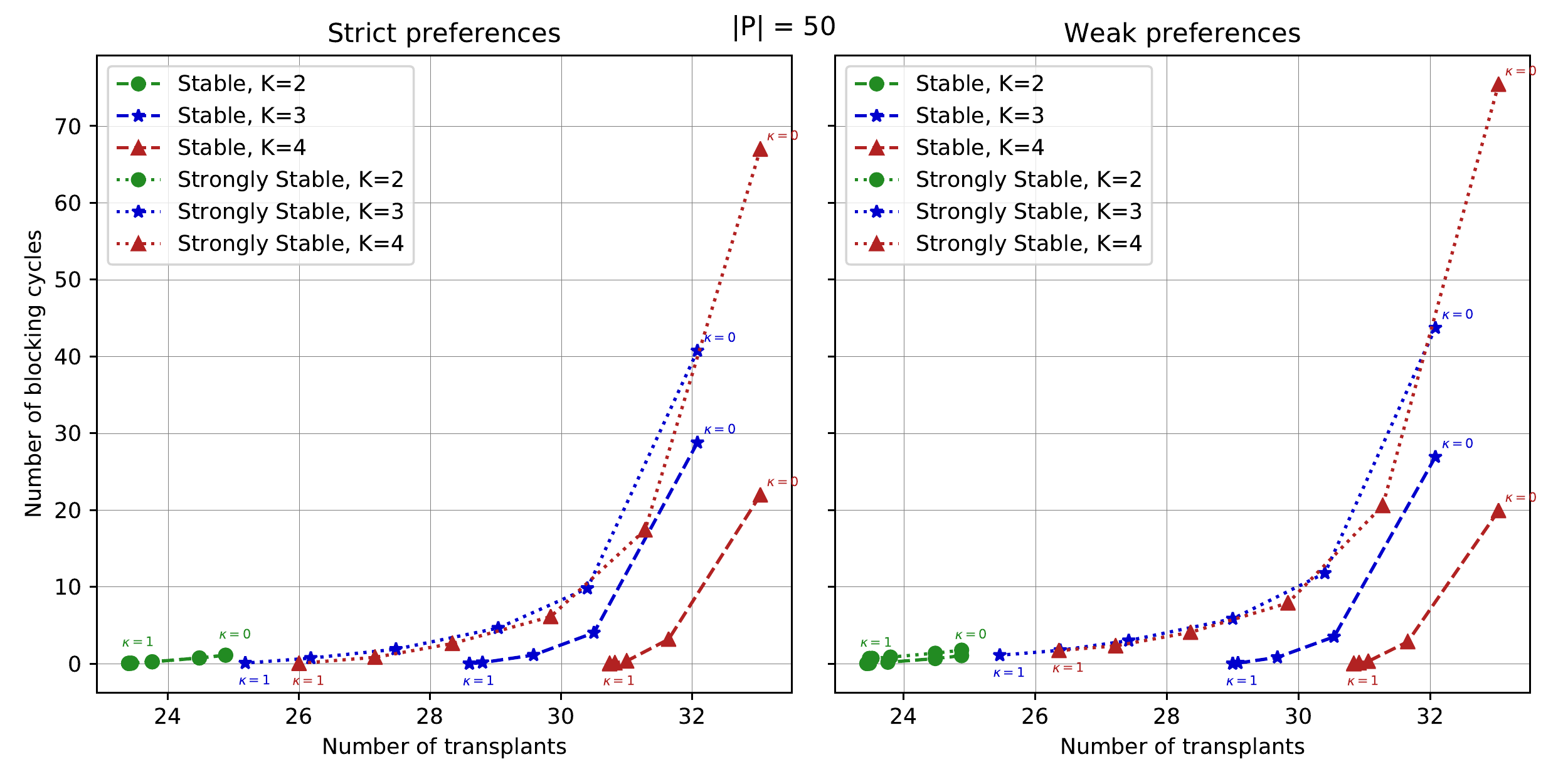}
\caption{Trade-off between stability  and  maximum   number of transplants  for instances with $|P| = 50$;  $\kappa = 0, 0.05,0.1, 0.15, 0.2, 1$.}\label{fig:50ps}
\end{figure}

\begin{figure}[htbp]
\includegraphics[scale = 0.6]{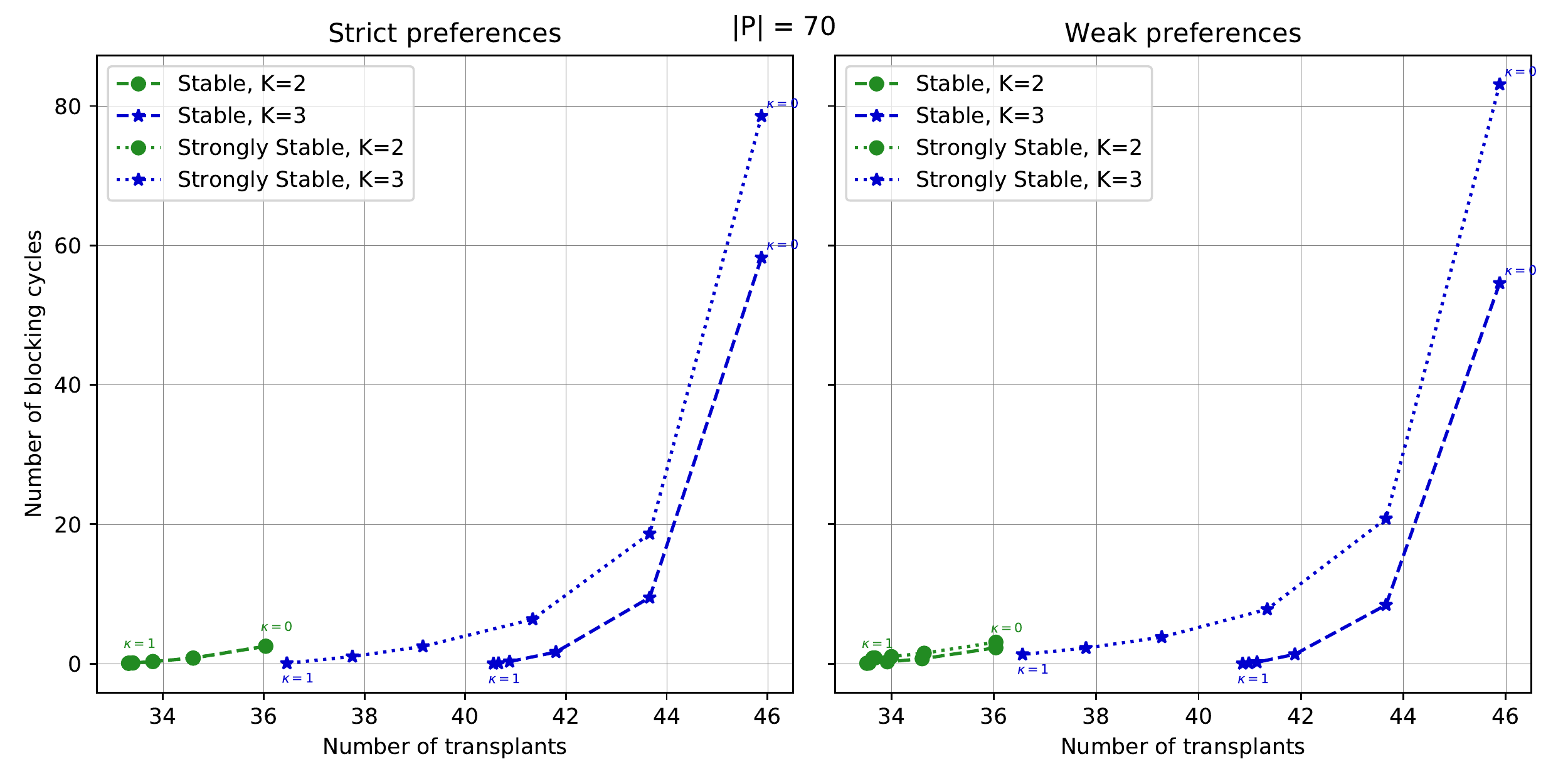}
\caption{Trade-off between stability and  maximum  number of transplants  for instances with $|P| = 70$; $\kappa = 0.05,0.1, 0,15, 0.2$.}\label{fig:70ps}
\end{figure}

\begin{figure}[htbp]
\includegraphics[scale = 0.6]{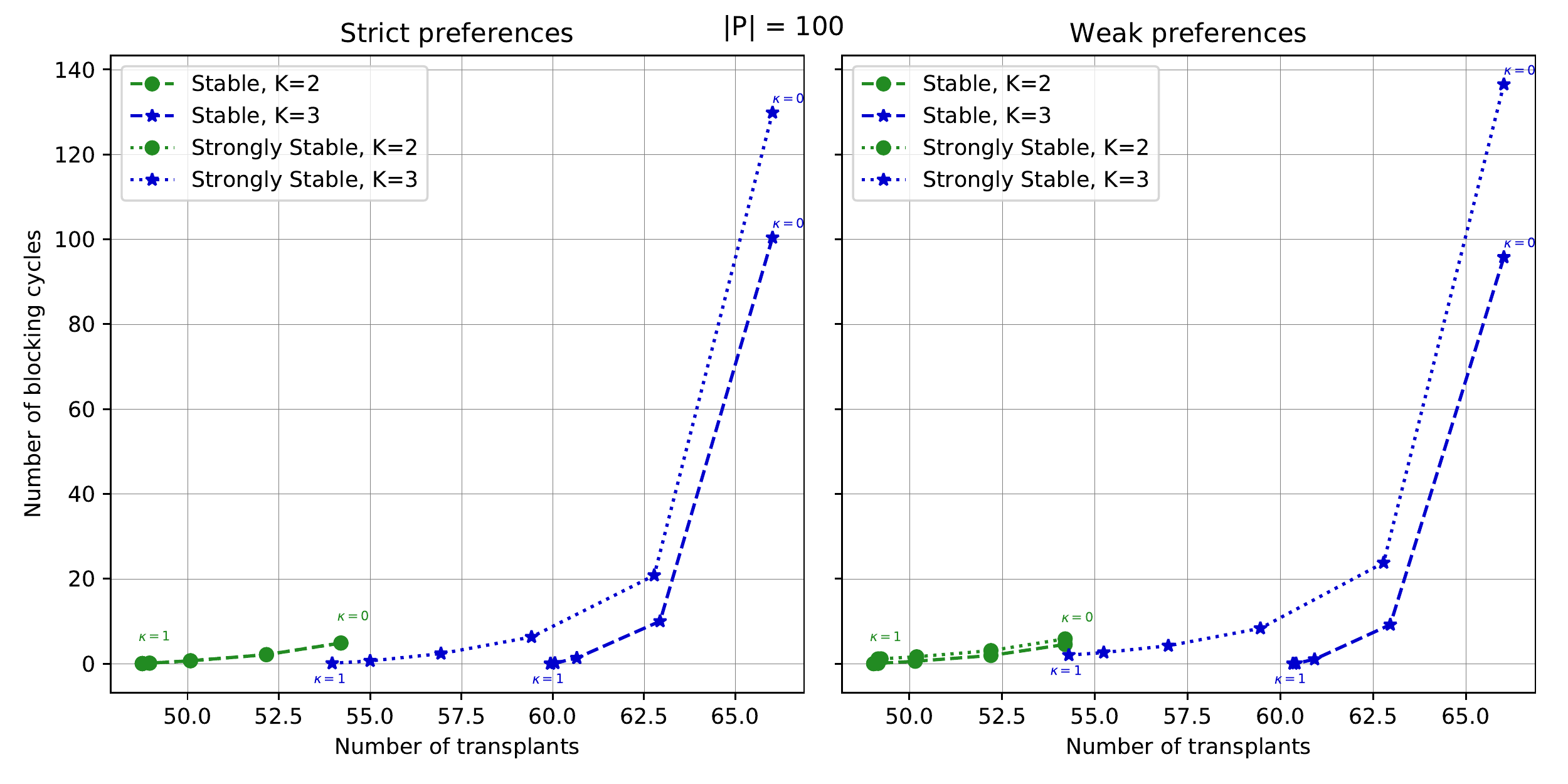}
\caption{Trade-off between stability and  maximum   number of transplants  for instances with $|P| = 100$;
$\kappa = 0, 0.05,0.1, 0.15, 0.2, 1$.}\label{fig:100ps}
\end{figure}

In Figures~\ref{fig:20ps}-\ref{fig:100ps} we plot the two objective function values for stable and strongly stable exchanges and for different values of $K$, representing those values separately for instances with strict (left hand-side graphs) and weak (right hand-side graphs) preferences.  Each figure presents average results for instances  of size $|P|$ = 20,50,70 and 100. Note that, since there were instances where strongly stable exchanges did not exist,  for $\kappa = 0$ the average number of blocking cycles is not always 0. 

Results clearly show that when $K = 2$, and when compared with larger values of $K$, the impact of stability requirements on the maximum number of transplants is low for smaller instances. However, when increasing the size of the instances, in particular for $|P| = 100$, stability affects the number of transplants with more than 10\% decrease, a value that is no longer negligible. In such a case, with an average number of blocking cycles of 5, the question is whether it is desirable to have at least 5 patients being transplanted from a donor that is not the most favorable for them or, instead, having a reduction of 5 to 6 transplants to guarantee that stability is met.
A similar conclusion can be drawn for larger values of $K$, in particular when strong stability is required and $K = 4$, as for this case the trade-off between the criteria in the two extreme solutions is high.

As reflected in the curves plotted for $K=3 \mbox{ and } 4$, strong stability requirements have a much stronger (negative) impact on the number of transplants, with  curve  \emph{Stable} always dominating curve \emph{Strongly Stable}. Hence, as expected, strong stability is achieved with more sacrifice on the number of transplants and is aligned with the results in Tables~\ref{tab:strict} and~\ref{tab:weak} on the values  of the  price of stability ($ps.$).

Also noteworthy is that for $|P| \geq 50$ trade-offs significantly differ for $\kappa \geq 0.05$ and for $\kappa < 0.05$. This can be confirmed by the different slopes in the graph: in between  $\kappa$ = 0.05 and 1 a smaller reduction in the number of blocking cycles is reflected in a comparatively larger reduction in the number of transplants; while for $\kappa < 0.05$ solution stability (measured by the number of blocking cycles) is severely affected for a relatively small gain in the number of transplants.
For small instances  with $|P| = 20$ (see figure~\ref{fig:20ps}) the number of blocking cycles decreases nearly linearly for values of $\kappa = 0 \mbox{ to } 0.2$.

Interestingly, by comparing the result for $K=3$ and $K=4$ for instances with $|P| = 20, 50$ we  observe that when we consider stability (curve \emph{Stable}) the number of blocking cycles does not differ significantly for different values of $\kappa$, i.e. one can slightly increase the number of transplants considering larger values of $K$ with no sacrifice in the number of blocking cycles. However, for the case of strong stability, for smaller values of $\kappa$ ($\kappa = 0 \mbox{ and } 0.05$) the number of blocking cycles for $K=4$ is in some cases nearly twice as large as for $K=3$. At the same time, for strongly stable solutions ($\kappa = 0$) the number of transplants does not differ significantly for $K=3,4$.
Taking into account that the problem difficulty increases dramatically when switching from $K=3$ to 4, one may consider that for strong stability the increase of $K$ is not primordial.

\par

\section{Conclusions}\label{sec:conclusion}

In this paper we consider kidney exchange programs in which patients may have preferences over potential donors, and where the aim is to find a maximum cardinality stable exchange. This is a problem of practical relevance when patients can be transplanted from several donors but some fit better medically than others.

We have advanced on the current state of the art by proposing three Integer Programming formulations capable of finding stable and strongly stable solutions, considering both strict and weak preferences. Furthermore, we have studied the impact of stability on the maximum number of transplants achievable by a pool. For doing so, we considered a bi-objective problem and studied the trade-off between (relaxed) stability, measured by the number of blocking cycles in a solution, and the maximum number of transplants. Such analysis is again of practical relevance: it can provide a solution with minimum number of blocking cycles for the decision maker that does not reduce the number of transplants below a certain limit.

Results show that the  cycle-edge formulation proposed in this paper retains
advantages from both the cycle and the edge formulation is more efficient than the two other formulations at solving the problems. This becomes even more evident when $K = 4$. Results also show that finding strongly stable exchanges is slightly easier than finding stable exchanges: all formulations need shorter CPU
times and more instances are solved within the same time limit.

In what concerns relaxation of stability requirements results show that when $K = 2$, and if compared with larger values of $K$, the impact of
stability requirements on the maximum number of transplants is low for smaller instances. However, when problem size increases, such value is no longer negligible. This conclusion is also valid for bigger values of $K$.  Results also show that strong stability is achieved with more sacrifice on the
number of transplants than stability.

\bibliographystyle{plain}
\bibliography{KEPstable}

\end{document}